\documentclass[12pt]{article}
\usepackage{graphicx}
\usepackage{amsmath}

\oddsidemargin  - 5 mm
\evensidemargin - 5 mm

\input{tcilatex}

\begin{document}

\date{}
\title{Algebraic Geometry Approach in Theories with Extra Dimensions I.
Application of Lobachevsky Geometry}
\author{Bogdan G. Dimitrov \thanks{%
Electronic mail: bogdan@theor.jinr.ru} \\
Bogoliubov Laboratory for Theoretical Physics\\
Joint Institute for Nuclear Research \\
6 Joliot - Curie str. \\
Dubna 141980, Russia}
\maketitle

\begin{abstract}
\ \ This present paper has the purpose to find certain physical appications
of Lobachevsky geometry and of the algebraic geometry approach in theories
with extra dimensions. It has been shown how the periodic properties of the
uniformization functions-solutions of cubic algebraic equations in gravity
theory enable the orbifold periodic identification of the points $\pi r_{c}$
and $-\pi r_{c}$ under compactification. It has been speculated that
corrections to the extradimensional volume in theories with extra dimensions
should be taken into account due to the non-euclidean nature of the
Lobachevsky space. It has been demonstrated that in the Higgs mass
generation model with two branes (a \textquotedblright
hidden\textquotedblright\ and a \textquotedblright
visible\textquotedblright\ one), to any mass on the visible brane there
could correspond a number of physical masses. Algebraic equations for $4D$
Schwarzschild Black Holes in higher dimensional brane worlds have been
obtained.
\end{abstract}

\vskip .5cm

\section{\protect\bigskip INTRODUCTION}

\bigskip It is commonly accepted in gravity theory [1] that the
contravariant metric tensor $g^{ij}$ is at the same time an inverse one to
the covariant one, i.e. 
\begin{equation}
g_{ij}g^{ik}=\delta _{i}^{k}\text{ \ \ \ .}  \tag{1.1}
\end{equation}%
From a more general point of view, the above equality sets up a
correspondence between the covariant and contravariant metric tensor
components, due to which the contravariant components cannot be considered
as independent from the covariant ones.

However, in more general theories of gravity [2,3,4] - theories with
covariant and contravariant metrics and affine connections, the covariant
metric tensor components are treated independently from the contravariant
ones (further they shall be denoted with the "tilda" sign), which means that 
\begin{equation}
g_{ij}\widetilde{g}^{ik}=l_{i}^{k}(\mathbf{x})\text{ \ \ \ ; \ \ \ \ }e_{i}%
\widetilde{e}^{k}=l_{i}^{k}(\mathbf{x})\text{ \ \ \ ,}  \tag{1.2}
\end{equation}%
where $l_{i}^{k}(\mathbf{x})$ are some (tensor) functions of the space -
time coordinates. Consequently, there is no longer a correspondence between
the covariant and contravariant metric tensor components, which should be
treated within the framework of the affine geometry approach [5,6,7].
Unfortunately, the components of the function $l_{j}^{i}(x)$ cannot be
determined from any physical considerations. That is why in [8] it was
proposed to find these functions by requiring the gravitational Lagrangian
with the more generally defined contravariant metric components to be the
same as the Lagrangian in the standard gravitational theory (i.e. with the
usual inverse contravariant tensor $g^{ij}$). As a result, a multivariable
cubic algebraic equation was obtained (see Appendix A) for the choice of the
contravariant metric tensor $\widetilde{g}^{ij}$ in the form of the
factorized product $\widetilde{g}^{ij}=dX^{i}dX^{j}$.

It should be remembered also that the choice of the contravariant metric
components in the known gravity theory as inverse to the covariant
components is a \textit{mathematical convention} - there is not a single
gravitational experiment, \textit{measuring directly} the contravariant
metric components. \textit{Consequently one may formulate the problem are
there physical reasons and considerations why this should be so. }If no such
straightforward physical reasons can be formulated , one has the right to
investigate what could be the physical consequences from such a more general
assumption.

The two parts of this paper will be related to the application of the
algebraic geometry approach [9, 10] and of Lobachevsky geometry in theories
with extra dimensions. The first part will deal mostly with the application
of Lobachevsky geometry, and the problem about Higgs mass creation in
theories with visible and hidden branes will be considered in the framework
of the theories with covariant and contravariant metrics. The visible and
hidden branes are situated at the orbifold fixed points $\Phi =0$ and $\Phi
=\pi $ and the covariant components of the visible brane are determined as 
\begin{equation}
g_{\mu \nu }^{vis}(X^{\mu })=G_{\mu \nu }(X^{\mu },\Phi =\pi
)=e^{-2kr_{-}\pi }g_{\mu \nu }\text{ \ . \ \ \ \ \ }  \tag{1.3}
\end{equation}%
The first problem, raised in this paper is: if in a more general
gravitational theory the contravariant components are not determined as the
inverse to the covariant ones, then their "scaling" in the action for the
fundamental Higgs field 
\begin{equation}
S_{vis}=\dint d^{4}X\sqrt{-g}_{vis}\left[ g_{vis}^{\mu \nu }D_{\mu
}H^{+}D_{\nu }H-\lambda \left( \mid H\mid ^{2}-v_{0}^{2}\right) ^{2}\right] 
\text{ \ \ \ }  \tag{1.4}
\end{equation}%
will not be as $g^{\mu \nu (vis)}=e^{2kr_{-}\pi }g^{\mu \nu }$.
Consequently, since the normalization of the fields determines the physical
masses, to any mass $m_{0}$ on the visible three-brane will no longer
correspond a single physical mass $m$ according to the relation $%
m=e^{-kr_{-}\pi }m_{0}$, but instead a multitude of physical masses.

The second problem, again related to the algebraic geometry approach, is the
orbifold identification of the points $\pi r_{C}$ and $-\pi r_{C}$ under
orbifold compactification. This is possible due to the previously
established in [10] property - the algebraic solutions of the cubic
multivariable equation for reparametrization invariance of the gravitational
Lagrangian represent uniformization functions, depending on the periodic
complex coordinate $z$.

The rest of the problems in this paper deal with the application of
Lobachevsky geometry in theories with extra dimensions. The basic fact in
these theories is the following one - the fundamental scale of gravity $%
M_{pl}$ is related to the gravity scale $M_{fund}$ in the $(4+d)-$%
dimensional space as 
\begin{equation}
M_{pl}^{2}=M_{fund}^{2+d}r_{i}^{d}=M_{fund}^{2+d}\text{ }V^{d}\text{ \ \ \ .}
\tag{1.5}
\end{equation}%
It is very important to realize that this relation is obtained on the base
of the factorizing approximation for the effective action in the form 
\begin{equation}
S_{eff}=\int d^{4}X\dint\limits_{0}^{\pi r_{-}}dy2M^{3}r_{c}e^{-2\widetilde{k%
}\mid y\mid }\sqrt{g}R\text{ \ \ \ ,}  \tag{1.6}
\end{equation}%
where the $5D$- metric is chosen with an exponentially suppressed
\textquotedblright warp\textquotedblright\ factor in front of the flat $4D$
Minkowski \ metric ($0\leq y\leq \pi r_{c}$) 
\begin{equation}
ds^{2}=e^{-2\widetilde{k}\mid y\mid }\eta _{\mu \nu }dX^{\mu }dX^{\nu
}+dy^{2}\text{ \ \ .}  \tag{1.7}
\end{equation}%
So the third problem, solved in this paper is: if another coordinate
transformation is applied, for example a one, related to the distance $\rho $
in the Lobachevsky geometry, then what happens with the extradimensional
coordinate and is the factorization approximation valid? The last means -
can the whole $5D$ spacetime be represented as a direct product of two
spaces? However, yet from the early investigations on Lobachevsky geometry
it was known that power - like corrections in $\frac{r}{c}$ ($r$- the
Euclidean distance) arise to the volume element, which in view of the
relation (1.5) will result also in corrections to the gravitational
couplings.

In fact, the cornerstone and basic reasoning for the application of the
Lobachevsky geometry in theories with extra dimensions is the similarity of
the $5D$ metric (1.7) with the $3D$ metric of a spacetime with a constant
negative curvature $ds^{2}=d\rho ^{2}+e^{-\frac{2\rho }{R}}d\sigma ^{2}$.
Although this metric is a three-dimensional one, multidimensional
Lobachevsky spaces have also been investigated long time ago in the
monographs of Rosenfel'd [11, 12, 13]. Now, if one makes the analogy between
the two metrics, another important question arises - what is the origin of
the constant $\widetilde{k}$ in the metric (1.7)? Unfortunately, this is not
clarified in the existing literature.In this paper it is shown that for the
choice $\widetilde{k}\neq \frac{1}{c}$, where $c$ is the Lobachevsky
constant, there will be an exponential increase of the extradimensional
distance (along the $y$ coordinate). This can be attributed to the
non-euclidean nature of the geometry. One more confirmation of the
non-euclidean geometry - if the limit $\widetilde{k}=\frac{1}{c}\rightarrow
0 $ is assumed (which means that $c\rightarrow \infty $ and then the
euclidean geometry is recovered), then it is impossible to set up the
fine-tuning $kr_{C}\approx 50$. The last is necessary if one would like to
have a five-dimensional Plack mass, not very far from the electroweak scale $%
M_{W}\approx TeV$. Consequently, the non-euclidean effects of the geometry
really come into play, and this is the fourth problem, discussed in this
paper. The fifth problem concerns the Riemann scalar curvature invariant $%
R_{ABCD}R^{ABCD}$, which has a singularity at $r\rightarrow 0$ and at $\mid
y\mid \rightarrow \infty $. Physically this may mean that there might be a
non - vanishing energy flow into the bulk singularities. So the problem now
may be formulated as follows: will the singularities remain or vanish in
gravitational theories with covariant and contravariant metrics? In other
words, the question is whether the scalar curvature invariant can be
preserved in such theories, which would signify the presence of the same
type of singularities. A fourth - order algebraic equation has been obtained
for the preservation of this invariant.

Appendix A contains a brief review of the algebraic geometry approach in
gravity theory, some aspects of which were used in section 7.

\section{\protect\bigskip FUNDAMENTAL \ PARALELLOGRAM \ ON \ THE \ COMPLEX \
PLANE, ORBIFOLD \ COMPACTIFICATION \ AND \ PERIODIC \ IDENTIFICATION}

\bigskip As it is well - known, a class of two - dimensional metrics exists
[14] 
\begin{equation}
ds^{2}=R^{2}\frac{(a^{2}-v^{2})du^{2}+2uvdudv+(a^{2}-u^{2})dv^{2}}{%
(a^{2}-u^{2}-v^{2})^{2}}\text{ \ \ \ ,}  \tag{2.1}
\end{equation}%
representing the linear element of a unit surface in the Lobachevsky space
with a constant negative curvature $-\frac{1}{R^{2}}$. Performing the
transformations 
\begin{equation}
\frac{a^{2}-u\text{ }u_{0}-vv_{0}}{\sqrt{a^{2}-u^{2}-v^{2}}}=ae^{-\frac{\rho 
}{R}}\text{ \ \ \ \ \ ;\ \ \ }\frac{u_{0}v-uv_{0}}{a^{2}-u\text{ }%
u_{0}-vv_{0}}=\frac{\sigma }{R}\text{ \ \ \ \ ,}  \tag{2.2}
\end{equation}%
the above metric (2.1) can be rewritten as 
\begin{equation}
ds^{2}=d\rho ^{2}+e^{-\frac{2\rho }{R}}d\sigma ^{2}\text{ \ \ \ \ ,} 
\tag{2.3}
\end{equation}%
which turns out to be similar to the metric 
\begin{equation}
ds^{2}=e^{-2kr_{-}\Phi }\eta _{\mu \nu }dx^{\mu }dx^{\nu }+r_{c}^{2}d\Phi
^{2}\text{ \ \ \ \ \ \ ,}  \tag{2.4}
\end{equation}%
extensively used in the first version of the Randall - Sundrum model [15].
In (2.4) $\eta _{\mu \nu }$ is the flat Minkowski metric, $0\leq \Phi \leq
\pi $ and the extra dimension is a finite interval, whose size is set by the
compactification radius $r_{c}$. A nice and effective generalization of this
model implies that the SM (Standard Model) particles and forces with the
exception of gravity are confined to a 4 - dimensional subspace, but within
a $(4+n)$- dimensional spacetime.

Let us now remind that the complex coordinate $z$ is the argument of the
Weierstrass function and as mentioned, it is defined on the lattice $\Lambda
=\{m\omega _{1}+n\omega _{2}\mid m,n\in Z;$ $\omega _{1},\omega _{2}\in C,Im%
\frac{\omega _{1}}{\omega _{2}}>0\}$ on the two - dimensional projective
plane $CP^{2}$. Then let us define the complex uniformization coordinate $z$
as $z=\pi r_{c}(\cos \Phi +i\sin \Phi )$ and $0\leq \Phi \leq \pi $ is the
periodic coordinate. Under the transformation $\Phi =arctg\frac{z}{r_{c}}$,
the metric (2.4) will transform as 
\begin{equation}
ds^{2}=e^{-2kr_{-}\frac{r_{c}}{\sqrt{z^{2}+r_{c}^{2}}}}\eta _{\mu \nu
}dx^{\mu }dx^{\nu }+\frac{r_{c}^{4}}{\sqrt{z^{2}+r_{c}^{2}}}dz^{2}\text{ \ \
\ \ \ .}  \tag{2.5}
\end{equation}%
Now the advantage of such a formulation is clear: the nice properties of the
Weierstrass function and its derivative 
\begin{equation}
\rho ^{^{\prime }}(z+\omega _{i})=\rho ^{^{\prime }}(z)\text{ \ \ \ ; \ \ \
\ }\rho (\pi r_{c})=\rho (-\pi r_{c})\text{ \ \ \ \ \ }  \tag{2.6}
\end{equation}%
exactly matches the requirement for orbifold identification of the points $%
+\pi r_{c}$ and $-\pi r_{c}$. In other words, by making the above
transformation for the periodical coordinate $\Phi $\ of the additional
extra dimension [15, 16], a periodical identification is achieved of the
identical points under orbifold compactification with a fundamental domain
of length $2\pi r_{c}$\ with the lattice points of the fundamental
paralellogram on the complex plane. A general overview of orbifold
compactifications is presented in [17].

\section{\protect\bigskip FACTORIZATION \ AND \ NON - FACTORIZATION \ OF \
THE \ VOLUME \ ELEMENT \ }

Obtaining some estimates for the fundamental length $2\pi r_{c}$ would be
interesting, since for $d$ additional compactified dimensions, each one of
radius $r_{i}$, the fundamental (Planck) scale of gravity is related to the
gravity scale in the $(4+d)$- dimensional space as [18, 19, 20] 
\begin{equation}
M_{pl}^{2}=M_{fund}^{2+d}r_{i}^{d}=M_{fund}^{2+d}\text{ }V^{d}\text{ \ \ \ .}
\tag{3.1}
\end{equation}

The estimate of the volume of the extradimensional space is important, since
by taking a large volume the large discrepancy between the Planck scale of $%
10^{19}$\bigskip GeV and the electroweak scale of $100$ GeV can be
diminished and thus the hierarchy problem can be solved. For a derivation of
the relation (3. 1) between the gravity scales on the base of dimensional
analysis of the higher - dimensional Einstein - Hilbert action, one may use
the review article [21]. In such a case, the metric (2.4) should be
generalized to the $(d+4)$- dimensional metric of the Lobachevsky space.
Naturally, the most simple case [21] is of a flat extradimensional space,
when $V^{d}=(2\pi r$ $)^{d}$ and also a flat $4D$ Minkowski \ metric.
However, it would be much more interesting to consider a $(d+4)$ -
dimensional $ADS$ (Lobachevsky) space with a constant negative curvature,
whose volume element is given by [22] 
\begin{equation}
dV_{n}=\frac{c_{n}dx_{1}dx_{2}...dx_{n}}{(c^{2}-x_{\alpha }x_{\alpha })^{%
\frac{(n+1)}{2}}}\text{ \ \ \ }  \tag{3.2}
\end{equation}%
and can be found by splitting up the $n$- dimensional volume by means of $%
(n-1)$- dimensional hyperplanes, perpendicular to the coordinate axis.
Details can be found again in the monograph [22]. For example, the five- and
four- dimensional volume elements are calculated to be 
\begin{equation}
V_{5}=\frac{1}{12}\pi ^{2}c^{6}(sh\frac{4r}{c}-8sh\frac{2r}{c}+12\frac{r}{c})%
\text{ \ \ \ ,}  \tag{3.3}
\end{equation}%
\begin{equation}
V_{4}=\pi c^{3}(sh\frac{2r}{c}-\frac{2r}{c})\text{ \ \ \ \ \ ,}  \tag{3.4}
\end{equation}%
where $r$ denotes the natural (euclidean) length and $c=\frac{1}{k}$ is the
Lobachevsky constant - the unit length parameter for the Lobachevsky space,
which enters the expressions (3.3 - 3.4). Our purpose will be to see whether
the constant $k$ in the the exponential factor $e^{-2kr_{-}\Phi }$ in the $%
(d+4)$- dimensional analogue of the metric (2.4) can be identified with the
inverse power of the Lobachevsky constant ($k=\frac{1}{c}$). Unfortunately,
in most of the existing papers on theories with extra dimensions, the origin
and meaning of the parameter $k$ is not clarified.

\bigskip Since in the limit $c\rightarrow \infty $ the usual Euclidean
geometry is recovered [23], then the above formulaes would give the volumes
of the five- and of the four- dimensional (Euclidean) spheres respectively 
\begin{equation}
V_{5}=\frac{8}{15}\pi ^{2}r^{5}\text{ \ \ \ ; \ \ \ }V_{4}=\frac{4}{3}\pi
r^{3}(1+\frac{1}{5}\frac{r^{2}}{c^{2}}+...)\text{ }=\text{\ }\frac{4}{3}\pi
r^{3}\text{\ \ \ .}  \tag{3.5}
\end{equation}%
The volumes of the $n$- dimensional (Euclidean) spheres for $n=2\lambda $
and $n=2\lambda +1$ [22] 
\begin{equation}
V_{2\lambda }=\frac{\pi ^{\lambda }}{\lambda !}r^{2\lambda }\text{ \ \ \ \ ;
\ \ \ \ }V_{2\lambda +1}=\frac{2^{\lambda +1}\pi ^{\lambda }}{(2\lambda
+1)(2\lambda -1)....3.1}r^{2\lambda +1}\text{ }  \tag{3.6}
\end{equation}%
can also be derived in the limit $c\rightarrow \infty $ from the (recurrent)
formulae for the $n$- dimensional hyperbolic volume 
\begin{equation}
V_{n}=\frac{2\pi c^{2}}{(n-1)}\left[ \frac{P_{n-4}}{(n-2)}c^{n-2}sh^{n-2}%
\frac{r}{c}ch\frac{r}{c}-V_{n-2}\right] \text{ \ \ \ ,}  \tag{3.7}
\end{equation}%
where 
\begin{equation}
P_{2\lambda }=\frac{2\pi ^{\lambda +1}}{\lambda !}\text{ \ \ \ \ ; \ \ }%
P_{2\lambda +1}=\frac{2^{\lambda +2}\pi ^{\lambda +1}}{(2\lambda
+1)(2\lambda -1)(2\lambda -3)....3.1}\text{\ \ \ \ .}  \tag{3.8}
\end{equation}%
Therefore, only in the flat (Euclidean) $(4+d)$- dimensional space, which is
a product of a $4$- dimensional Minkowski space ($\mu ,\nu =1,2,..,4$) and a
flat $d$- (extra)dimensional space [21] 
\begin{equation}
ds^{2}=\eta _{\mu \nu }dX^{\mu }dX^{\nu }-r^{2}d\Omega _{(d)}^{2}\text{ \ \
\ ,}  \tag{3.9}
\end{equation}%
one can factorize the volume element in the $(4+d)$ Einstein - Hilbert
action (assuming also that $R^{(4+d)}=R^{(4)}$) as 
\begin{equation}
S_{4+d}=-M_{\ast }^{d+2}\int d^{4+d}X\sqrt{g^{(4+d)}}R^{(4+d)}=-M_{\ast
}^{(d+2)}\int d\Omega _{(d)}r^{d}\int d^{4}X\sqrt{g^{(4)}}R^{(4)}\text{ .} 
\tag{3.10}
\end{equation}%
It is clear however from expressions (3.3 - 3.4), (3.7) that in the general
case of a multidimensional non-euclidean (Lobachevsky) space such a
factorization of the volume element is impossible. Even in the limit of
small ratios $\frac{r}{c}$, possible corrections to the volume element (see
f.(3. 5)) have to be taken into account and therefore, the non-euclidean
geometry would \textquotedblright induce\textquotedblright\ correction terms
in the relation between the gravitational couplings.

\section{\protect\bigskip COORDINATE \ TRANSFORMATIONS \ WITH \ THE \
LOBACHEVSKY \ CONSTANT \ IN \ "WARP" \ TYPE \ OF \ METRICS}

Let us turn to the other frequently used case [24, 25] of a $5D$- metric
with an exponentially suppressed \textquotedblright warp\textquotedblright\
factor in front of the flat $4D$ Minkowski \ metric ($0\leq y\leq \pi r_{c}$%
) 
\begin{equation}
ds^{2}=e^{-2\widetilde{k}\mid y\mid }\eta _{\mu \nu }dX^{\mu }dX^{\nu
}+dy^{2}  \tag{4.1}
\end{equation}%
and for the moment it shall be assumed that $\widetilde{k}$ is different
from the constant $k=\frac{1}{c}$. The five - dimensional effective action
can be factorized as [24] 
\begin{equation}
S_{eff}=\int d^{4}X\dint\limits_{0}^{\pi r_{-}}dy2M^{3}r_{c}e^{-2\widetilde{k%
}\mid y\mid }\sqrt{g}R\text{ \ \ \ }  \tag{4.2}
\end{equation}%
and the metric (4.1) is chosen so that the $5$- dimensional Ricci scalar
curvature is equal to the $4$- dimensional Minkowski one. From (4.2), the
matching relation between the gravitational couplings is obtained to be [24] 
\begin{equation}
M_{pl}^{2}=2M^{3}\dint\limits_{0}^{\pi r_{-}}dye^{-2\widetilde{k}\mid y\mid
}=\frac{M^{3}}{\widetilde{k}}(1-e^{-2\widetilde{k}r_{-}\pi })\ \ \ . 
\tag{4.3}
\end{equation}%
\ \ $\ \ \ \ $

Let us note that the result of the integration in (4.2 - 4.3) will not be
coordinate independent. In other words, if we map the $4D$ Minkowski part of
the metric with an exponentially \textquotedblright
damped\textquotedblright\ prefactor into a flat Minkowski $4D$ metric
without the exponential prefactor, this would result in the appearence of an
exponentially growing prefactor in front of the $dy^{2}$ part of the metric
(4.1). To illustrate this, let us use a coordinate transformation, similar
to the one, used in [26] 
\begin{equation}
X^{1}=a\text{ }ch\frac{\rho }{c}\text{ \ \ ; \ \ \ }X^{2}=b\text{ }sh\frac{%
\rho }{c}\sin \Theta \cos \varphi \text{ \ \ \ ,}  \tag{4.4}
\end{equation}%
\begin{equation}
X^{3}=b\text{ }sh\frac{\rho }{c}\sin \Theta \sin \varphi \text{ \ \ ; \ \ \ }%
X^{4}=b\text{ }sh\frac{\rho }{c}\cos \Theta \text{ \ \ \ .\ }  \tag{4.5}
\end{equation}%
The signature of the Minkowski space is $(+,-,-,-)$, i.e. $\eta _{\mu \nu
}=(+1,-1,-1,-1)$, $\rho $ is the distance in the Lobachevsky space, related
to the euclidean distance $r$ by the formulae 
\begin{equation}
r=c\text{ }sh\frac{\rho }{c}\text{ \ \ .}  \tag{4.6}
\end{equation}%
The constants $a$ and $b$ are to be chosen so that a flat Minkowski metric
without any prefactor is obtained. In fact, if $\widetilde{k}=\frac{1}{c}$,
the metric (4.1) will be exactly the one, known from Lobachevsky geometry.
Now we shall establish the physical meaning of the relation $\widetilde{k}=%
\frac{1}{c}$, but in reference to theories with extra dimensions. An
elementary introduction into Lobachevsky geometry can be found in [27] and a
more comprehensive and detailed exposition - in [28].

For the choice $a=b=ce^{\widetilde{k}\mid y\mid }$ and after applying the
transformations (4.4 - 4.5), the metric (4.1) can be rewritten as 
\begin{equation}
ds^{2}=-d\rho ^{2}-c^{2}sh^{2}\frac{\rho }{c}d\Theta ^{2}-c^{2}sh^{2}\frac{%
\rho }{c}\sin ^{2}\Theta d\varphi ^{2}+(c\widetilde{k})^{2}e^{2\widetilde{k}%
\mid y\mid }dy^{2}\text{ \ \ \ .}  \tag{4.7}
\end{equation}%
Obviously, the first three terms give the unit length element in the
Lobachevsky space, which in the limit $c\rightarrow \infty $ (then from
(4.6) \ $\rho \rightarrow r$) gives the usual euclidean length element $%
dr^{2}+r^{2}(d\Theta ^{2}+\sin ^{2}\Theta d\varphi ^{2})$ in spherical
coordinates $(r,\Theta ,\varphi )$. Most interesting is the last term in
(4.7) - it goes to infinity if $\mid y\mid \rightarrow \infty $ and in the
limit $c\rightarrow \infty $ (when $c\neq \frac{1}{\widetilde{k}}$), but it
tends to $1$ in the limit $c\rightarrow \infty $ and when $\widetilde{k}=%
\frac{1}{c}$, which is physically reasonable because a flat euclidean
geometry is obtained, as it should be. Thus the exponential increase of the
\textquotedblright extra - dimensional\textquotedblright\ distance, when $%
c\neq \frac{1}{\widetilde{k}}$, can be regarded as an effect of the
non-euclidean nature of space-time. Indeed, it is physically unacceptable to
take the limit $\widetilde{k}=\frac{1}{c}\rightarrow 0$, because if the five
- dimensional Planck mass is assumed to be not very far from the electroweak
scale $M_{W}\approx TeV$, then a fine- tuning $\widetilde{k}$ $r_{c}\approx
50$ is needed [18].

\section{\protect\bigskip ALGEBRAIC $\ $EQUATIONS $\ $IN $\ 4D$ \
SCHWARZSCHILD \ BLACK \ HOLES \ IN \ HIGHER \ DIMENSIONAL \ BRANE \ WORLDS}

\bigskip Now suppose that the metric (4.1) does not contain a flat Minkowski 
$4D$ space, but a $4D$ black hole instead 
\begin{equation}
ds^{2}=e^{-2\widetilde{k}\mid y\mid }g_{\mu \nu }dX^{\mu }dX^{\nu }+dy^{2}%
\text{ \ \ ,}  \tag{5.1}
\end{equation}%
where 
\begin{equation*}
g_{\mu \nu }dX^{\mu }dX^{\nu }=-(1-\frac{2M}{r})dt^{2}+(1-\frac{2M}{r}%
)^{-1}dr^{2}+
\end{equation*}%
\begin{equation}
+r^{2}(d\Theta ^{2}+\sin ^{2}\Theta d\varphi ^{2})\text{ \ \ \ .}  \tag{5.2}
\end{equation}%
For such a model [32, 33], if a negative tension brane is introduced at a
distance $y=l<\infty $, the five dimensional BH singularity will have a
finite size and a black tube will extend into the bulk, thus interpolating
between the two black holes.

A possible application of the formalism in this paper is related to the
Riemann scalar curvature invariant $R_{ABCD}R^{ABCD}$ [32], which for the
background metric (5.2) and using the conventional contravariant metric
components $g^{ij}$ is calculated to be [32] 
\begin{equation}
R_{ABCD}R^{ABCD}=40k^{4}+\frac{48M^{2}e^{4k\mid y\mid }}{r^{6}}\text{ \ \ \ .%
}  \tag{5.3}
\end{equation}%
This expression contains an important physical information - it diverges at
the black hole singularity at $r=0$ and also at the $ADS$ horizon at $\mid
y\mid \rightarrow \infty $. The elimination of this singularity (i. e.
giving it a finite size) is the main motivation for introducing the second,
negative tension brane at a distance $y=L$. But even in the case of a single
brane configuration the presence of a singularity is essential since there
might be a non- vanishing energy flow into the bulk singularities, which is
not desirable. Such an energy flow will exist if the limit [32] 
\begin{equation}
\lim_{\mid y\mid \rightarrow \infty }\sqrt{-g}J_{(\mu )}^{y}=\lim_{\mid
y\mid \rightarrow \infty }\sqrt{-g}T^{yN}K_{N}^{(\mu )}  \tag{5.4}
\end{equation}%
is non- zero, where $J_{(\mu )}^{y}$ and $T^{yN}$ are the current and the
energy- momentum tensor \ of a massless scalar field and $K_{N}^{(\mu
)}=e^{2A}\delta _{M}^{\mu }$ ($\mu =t,\Theta ,\varphi $) is the Killing
vector for the BH metric (5.1 - 5.2).

Therefore, it is essential to check whether the presence of the singularity
in (5.3) and of the zero energy flow in (5.4) will be confirmed if the same
scalar curvature $R$ will be obtained by contracting the Riemann tensor with
another contravariant metric tensor field $\widetilde{g}^{ij}$ such that 
\begin{equation}
R=g^{AC}g^{BD}R_{ABCD}=\widetilde{g}^{AC}\widetilde{g}^{BD}\widetilde{R}%
_{ABCD}  \tag{5.5}
\end{equation}%
where $\widetilde{R}_{ABCD}$ is the modified Riemann tensor\textbf{\ }with
the more generally defined contravariant metric tensor $\widetilde{g}^{ij}$%
\textbf{\ } 
\begin{equation*}
\widetilde{R}_{ABCD}\equiv \frac{1}{2}%
(g_{AD,BC}+g_{BC,AD}-g_{AC,BD}-g_{BD,AC})+g_{np}(\widetilde{\Gamma }_{BC}^{n}%
\widetilde{\Gamma }_{AD}^{p}-\widetilde{\Gamma }_{BD}^{n}\widetilde{\Gamma }%
_{AC}^{p})=
\end{equation*}%
\begin{equation}
=\frac{1}{2}(....)+g_{np}g_{rs}g_{qt}\widetilde{g}^{ns}\widetilde{g}%
^{pt}(\Gamma _{BC}^{n}\Gamma _{AD}^{p}-\Gamma _{BD}^{n}\Gamma _{AC}^{p})%
\text{ \ \ \ .}  \tag{5.6}
\end{equation}%
If the scalar curvature $R$, the connection $\Gamma _{AB}^{C}$ are
calculated from the initially given metric, equation (5.5) can be treated as
an fourth- order algebraic equation with respect to the components $%
\widetilde{g}^{AB}$ and as an eight - order\textbf{\ }algebraic equation
with respect to the variables $dX^{A}$, if again the factorization $%
\widetilde{g}^{AB}=dX^{A}dX^{B}$ is used. This example clearly shows the
necessity to go beyond the assumption about the contravariant metric
factorization. But on the other hand, even if the factorization assumption
is used, the same scalar curvature can be obtained by contracting the
(modified) Ricci tensor with the contravariant metric tensor $\widetilde{g}%
^{AB}$, i. e. $R=\widetilde{g}^{AB}R_{AB}$, which was in fact the cubic
algebraic equation, investigated in [8, 9, 10].

But there is also one more way for obtaining the scalar curvature $R$ - by
assuming that the following algebraic equation with the usual Riemann tensor
components holds 
\begin{equation}
R=\widetilde{g}^{AC}\widetilde{g}^{BD}R_{ABCD}\text{ \ \ .}  \tag{5.7}
\end{equation}%
Fortunately, this equation is second order with respect to $\widetilde{g}%
^{AB}$ and fourth order with respect to $dX^{A}$ and moreover, it does not
contain any derivatives of the components $\widetilde{g}^{AB}$ and $dX^{A}$.

Let us now assume that in the framework of the factorization assumption,
both equations (5.5) and (5.7) are fulfilled. Then the fulfillment of these
equations is a necessary condition for the preservation of the scalar
curvature invariant because 
\begin{equation}
R_{ABCD}R^{ABCD}=R_{ABCD}\widetilde{g}^{Ai}\widetilde{g}^{Bj}\widetilde{g}%
^{Ck}\widetilde{g}^{Dl}\widetilde{R}_{ijkl}=  \tag{5.8}
\end{equation}%
\begin{equation}
=\left( R_{ABCD}dX^{A}dX^{B}dX^{C}dX^{D}\right) \left( \widetilde{R}%
_{ijkl}dX^{i}dX^{j}dX^{k}dX^{l}\right) =  \tag{5.9}
\end{equation}%
\begin{equation}
=\left( R_{ABCD}\widetilde{g}^{AC}\widetilde{g}^{BD}\right) \left( 
\widetilde{R}_{ijkl}\widetilde{g}^{ik}\widetilde{g}^{jl}\right) =R^{2}\text{
\ \ \ \ .}  \tag{5.10}
\end{equation}%
Motivated by the necessity to investigate lower degree algebraic equations,
one may take equation (5.7) and also the equation 
\begin{equation}
\widetilde{g}^{AC}\widetilde{g}^{BD}\widetilde{R}_{ABCD}-\widetilde{g}^{AC}%
\widetilde{g}^{BD}R_{ABCD}=0\text{ \ \ \ \ .}  \tag{5.11}
\end{equation}%
A subclass of solutions of this equation will be represented by the
algebraic equation 
\begin{equation}
\widetilde{g}^{BD}\widetilde{R}_{ABCD}-\widetilde{g}^{BD}R_{ABCD}=0\text{ \
\ \ \ }  \tag{5.12}
\end{equation}%
(cubic with respect to $\widetilde{g}^{AB}$ and of sixth order with respect
to $dX^{A}$) and another, more restricted class of solutions - by the
equation 
\begin{equation}
\widetilde{R}_{ABCD}-R_{ABCD}=0\text{ \ \ \ \ ,}  \tag{5.13}
\end{equation}%
which is quadratic in $\widetilde{g}^{AB}$ and quartic with respect to $%
dX^{A}$. Therefore, even in such a complicated case, the investigation of
the intersection varieties of the two quartic equations (5.7) and (5.13),
written respectively as (again, it shall be used that $\widetilde{\Gamma }%
_{ij}^{k}=dX^{k}dX^{s}g_{rs}\Gamma _{ij}^{r}$) \textbf{\ } 
\begin{equation}
dX^{A}dX^{B}dX^{C}dX^{D}R_{ABCD}-R=0  \tag{5.14}
\end{equation}%
\textbf{\ }and\textbf{\ } 
\begin{equation*}
g_{np}g_{rs}g_{qt}(\Gamma _{BC}^{r}\Gamma _{AD}^{q}-\Gamma _{BD}^{r}\Gamma
_{AC}^{q})dX^{n}dX^{s}dX^{p}dX^{t}-
\end{equation*}%
\begin{equation}
-g_{np}(\Gamma _{BC}^{n}\Gamma _{AD}^{p}-\Gamma _{BD}^{n}\Gamma _{AC}^{p})=0%
\text{ \ \ \ \ ,}  \tag{5.15}
\end{equation}%
\textbf{\ }may give some solutions for the contravariant metric tensor
components $\widetilde{g}^{AB}=dX^{A}dX^{B}$, which will preserve both the
scalar curvature and the scalar curvature invariant. Respectively, if only
the scalar curvature $R$ is to be preserved, one may find the solutions of
the algebraic equation (5.7) and then substitute them in the expression for
the scalar curvature invariant $R_{ABCD}R^{ABCD}$.

It is clear also that if one takes only equation (5.5) $R=\widetilde{g}^{AC}%
\widetilde{g}^{BD}\widetilde{R}_{ABCD}$ and not equation (5.7), from (5.8) -
(5.10) one may obtain not the equality $R_{ABCD}R^{ABCD}=R^{2}$, but an
fourth- order algebraic equation with respect to $dX^{A}$ for the
preservation of the scalar curvature invariant 
\begin{equation}
R.R_{ABCD}dX^{A}dX^{B}dX^{C}dX^{D}-R_{ABCD}R^{ABCD}=0\text{ \ \ \ \ .} 
\tag{5.16}
\end{equation}%
But this is not the only possibility. One may take also only\textbf{\ }%
equation (5.7) $R=\widetilde{g}^{AC}\widetilde{g}^{BD}R_{ABCD}$\textbf{\ }%
and disregard equation (5.5). Then the resulting algebraic equation from
(5.8) - (5.10) will be 
\begin{equation*}
\frac{1}{2}(g_{AD,BC}+g_{BC,AD}-g_{AC,BD}-g_{BD,AC})dX^{A}dX^{B}dX^{C}dX^{D}+
\end{equation*}%
\begin{equation*}
+g_{np}g_{rs}g_{qt}(\Gamma _{BC}^{r}\Gamma _{AD}^{q}-\Gamma _{BD}^{r}\Gamma
_{AC}^{q})dX^{A}dX^{B}dX^{C}dX^{D}dX^{n}dX^{s}dX^{p}dX^{t}-
\end{equation*}%
\begin{equation}
-R_{ABCD}R^{ABCD}=0\text{ \ \ .}  \tag{5.17}
\end{equation}%
This equation is of eight order and due to the presence of the last scalar
curvature invariant term it is impossible to find subclasses of solutions of
(lower - order) algebraic equations, as in the case of eq. (5.11).

\section{\protect\bigskip COMPACTIFICATION \ RADIUS \ AND \ SCALAR \ FIELD \
EQUATION $\ $IN $\ 4D$ \ SCHWARZSCHILD \ BLACK \ HOLES \ IN \ HIGHER \
DIMENSIONAL \ BRANE \ WORLDS}

\textbf{\bigskip }In theories with extra dimensions, for example $(4+n)$-
dimensional Schwarzschild black hole [33, 34, 35, 36] 
\begin{equation}
ds^{2}=-h(r)dt^{2}+h^{-1}(r)dr^{2}+r^{2}d\Omega _{n+2}^{2}  \tag{6.1}
\end{equation}%
with 
\begin{equation}
h(r)=1-\left( \frac{r_{H}}{r}\right) ^{n+1}  \tag{6.2}
\end{equation}%
($r_{H}$- \ the horizon radius) it is important to distinguish between
distances $r\ll R_{1}$ ($R_{1}$- the compactification radius), when the BH
is a $(4+n)$- dimensional one, and distances $r\gg R_{1}$, when the BH
metric goes over to the usual four dimensional Schwarzschild metric 
\begin{equation}
ds^{2}=-(1-\frac{2M}{M_{Hr}^{2}})dt^{2}+(1-\frac{2M}{M_{Hr}^{2}}%
)^{-1}dr^{2}+r^{2}d\Omega ^{2}\text{ \ \ .}  \tag{6.3}
\end{equation}%
However, when solving the scalar wave equation $g^{IJ}\Phi _{I;J}=0$, there
is no way to introduce the scale factor $R_{1}$ in the solution of the
scalar equation. If this can be done, the scalar field behaviour can be
compared in the transition from one limit to another.

The use of the more general contravariant tensor $\widetilde{g}^{ij}$ gives
the opportunity to introduce such a scale factor. Let us first note that 
\begin{equation}
g_{AB}\widetilde{g}^{BC}=l_{A}^{C}(\mathbf{x})\text{ \ \ \ }\Rightarrow 
\text{ \ }\widetilde{g}^{BC}=l_{D}^{B}g^{DC}\text{ \ \ \ \ ,}  \tag{6.4}
\end{equation}
where $A,B,C,D$ concretely in this case will denote the $(4+n)$- dimensional
indices, $\mu ,\nu $- only the four- dimensional indices and $i,j,k$ denote
the indices of the additional $n$- dimensional space. Then the metric can be
represented as 
\begin{equation}
ds^{2}=g_{AB}dX^{A}dX^{B}=g_{\mu \nu }dX^{\mu }dX^{\nu
}+\sum_{i=5}^{n+4}l_{i}^{i}=ds_{(4)}^{2}+nR_{1}  \tag{6.5}
\end{equation}
where it has been assumed that $l_{i}^{i}=R_{1}$ for all $i$. Consequently,
some of the components $\widetilde{g}^{jB}$ of the contravariant metric
tensor can be expressed as 
\begin{equation}
\widetilde{g}^{jB}=l_{\nu }^{j}g^{\nu B}+l_{i}^{j}g^{iB}=l_{\nu }^{j}g^{\nu
B}+R_{1}g^{jB}+\underset{i\neq j}{l_{i}^{j}g^{iB}}  \tag{6.6}
\end{equation}
and evidently the solutions of the scalar wave equation will depend on the
compactification radius $R_{1}$.

\section{\protect\bigskip A \ COMPLIMENTARY \ PROPOSAL \ FOR \ HIGGS \ MASS
\ GENERATION \ IN \ THEORIES \ WITH \ TWO \ THREE - BRANES}

\bigskip Closely related to the above problem about the contravariant metric
tensor components as coupling constants is the problem about Higgs mass
generation in theories with two branes [15, 25] - the so called \textit{%
\textquotedblright hidden\textquotedblright }\ and \textit{%
\textquotedblright visible\textquotedblright }\ branes at the orbifold fixed
points $\Phi =0$ and $\Phi =\pi .$ The metric, which will be used is again
(2.4). These three branes couple to the four dimensional components $G_{\mu
\nu }$ of the bulk metric as [15] 
\begin{equation}
g_{\mu \nu }^{vis}(X^{\mu })=G_{\mu \nu }(X^{\mu },\Phi =\pi )\text{ \ \ \ \
; \ \ \ \ }g_{\mu \nu }^{hid}(X^{\mu })=G_{\mu \nu }(X^{\mu },\Phi =0)\text{%
\ .}  \tag{7.1}
\end{equation}%
The action includes the gravity part plus the action for the visible and
hidden branes and also the action for the fundamental Higgs field 
\begin{equation}
S_{vis}=\dint d^{4}X\sqrt{-g}_{vis}\left[ g_{vis}^{\mu \nu }D_{\mu
}H^{+}D_{\nu }H-\lambda \left( \mid H\mid ^{2}-v_{0}^{2}\right) ^{2}\right] 
\text{ \ \ \ ,}  \tag{7.2}
\end{equation}%
where $v_{0}$ is the vacuum expectation value (VEV) for the Higgs field $H$, 
$\lambda $ is a coupling constant [15]. Similar coupling of the
contravariant metric tensor components to a gauge field can be found also in
radion cosmology theories [37]. Since $g_{\mu \nu }^{vis}=e^{-2kr_{-}\pi
}g_{\mu \nu }$, it is believed that by a proper normalization of the fields
one can determine the physical masses. In particular, if the Higgs field
wave function is normalized as $H\rightarrow e^{kr_{-}\pi }H$, then 
\begin{equation}
S_{vis}=\int d^{4}X\sqrt{-g}\left[ g^{\mu \nu }D_{\mu }H^{+}D_{\nu
}H-\lambda \left( \mid H\mid ^{2}-e^{-2kr_{-}\pi }v_{0}^{2}\right) ^{2}%
\right] \text{ \ \ . }  \tag{7.3}
\end{equation}%
Therefore, since $v=e^{-2kr_{-}\pi }v_{0}$, any mass $m_{0}$\ on the visible
three- brane in the fundamental higher- dimensional theory will correspond
to a physical mass 
\begin{equation}
m=e^{-kr_{-}\pi }m_{0}\text{ \ \ \ ,}  \tag{7.4}
\end{equation}%
\ \textquotedblright measured\textquotedblright\ with the metric $g^{\mu \nu
}$\ in the effective Einstein- Hilbert action. If $kr_{c}\approx 50$\ (i.e. $%
e^{kr_{-}\pi }\approx 10^{15}$), this is the physical mechanism that is
supposed to produce TeV physical mass scales from mass parameters around the
Planck scale $\approx 10^{19}$\ GeV.

In the context of the developed approach in this paper, now it shall be
shown that the above physical mechanism of generation of TeV mass scales may
turn out to be more complicated and diverse. Namely, for a given scalar
curvature, there will be a multitude of contravariant metric tensors, thus
suggesting that the possibilities for the mass scales will be much more.

Following the earlier developed algebraic geometry approach in [9, 10],
which will be briefly reviewed also in Appendix A, the contravariant metric
tensor components $\widetilde{g}^{\mu \nu }$ can be written as 
\begin{equation}
\widetilde{g}^{\mu \nu }=dX^{\mu }dX^{\nu }=F_{\mu }(\mathbf{X}(z,v),\Phi
(z,v),z)F_{\nu }(\mathbf{X}(z,v),\Phi (z,v),z)\text{ \ .}  \tag{7.5}
\end{equation}%
It might seem strange that a \textit{particular choice} of the contravariant
metric components has been used. The important moment here is that for a
given metric and scalar curvature and no matter that the contravariant
metric components are \textit{not generally chosen}, \textit{there exist}
contravariant components, for which $g_{\alpha \mu }\widetilde{g}^{\mu \nu
}=l_{\alpha }^{\nu }\neq $ $\delta _{\alpha }^{\nu }$.

Further, the (contravariant) metric on the visible brane can be expressed as 
\begin{equation}
\widetilde{g}_{vis}^{\mu \nu }=L_{2}(z,v)\widetilde{g}^{\mu \nu }\text{ \ \ ,%
}  \tag{7.6}
\end{equation}%
where 
\begin{equation}
L_{2}(z,v)\equiv \frac{F_{\mu }(\mathbf{X}(z,v),\Phi (z,v)=\pi ,z)F_{\nu }(%
\mathbf{X}(z,v),\Phi (z,v)=\pi ,z)}{F_{\mu }(\mathbf{X}(z,v),\Phi
(z,v),z)F_{\nu }(\mathbf{X}(z,v),\Phi (z,v),z)}\text{ \ \ .}  \tag{7.7}
\end{equation}%
\textbf{\ }Formulaes (7.6) - (7.7) have been derived as a ratio of the
\textquotedblright visible\textquotedblright\ and the usual contravariant
metric components for each fixed indices $(\mu ,\nu )=(\mu _{1},\nu _{1})$
and without assuming that the points on the complex plane, for which \textbf{%
\ }$\Phi (z_{0},v_{0})=\pi $, are known. Further it shall be shown how the
calculation will be modified if these points are assumed to be known.

The transition from the four- dimensional variables $%
d^{4}X=dX_{1}dX_{2}dX_{3}dX_{4}$ to the two- dimensional complex variables $%
(z,v)$ can be performed by using the formulae 
\begin{equation}
d^{4}X=\sum_{1\leq i_{1}<i_{k}\leq 4}\det 
\begin{Vmatrix}
\frac{\partial X_{i_{1}}}{\partial z} & \frac{\partial X_{i_{k}}}{\partial v}
\\ 
\frac{\partial X_{i_{k}}}{\partial z} & \frac{\partial X_{i_{k}}}{\partial v}%
\end{Vmatrix}%
dz\wedge dv=L_{3}(z,v)dz\wedge dv\text{ \ \ \ ,}  \tag{7.8}
\end{equation}%
but since we are interested in rescaling only the Higgs field and the
contravariant metric as 
\begin{equation}
H\rightarrow \widetilde{H}f\text{ \ \ ;\ \ \ \ }v_{0}\rightarrow \widetilde{v%
}_{0}\text{ \ \ \ \ \ }(f-\text{a function})\text{ \ \ \ ,}  \tag{7.9}
\end{equation}%
the change of\textbf{\ }variables in the volume integration is not necessary
to be taken into account. Next it is necessary to find how the volume
element $\sqrt{-g}_{vis}$ of the visible brane can be expressed through the
volume element $\sqrt{-g}$ in terms of the metric (2.4). It can easily be
calculated that 
\begin{equation}
\sqrt{-g}=\sqrt{K_{1}(\Phi ,\frac{\partial \Phi }{\partial z},\frac{\partial
\Phi }{\partial v},\frac{\partial X^{\mu }}{\partial z},\frac{\partial
X^{\mu }}{\partial z})+e^{-4kr_{-}\Phi }K_{2}(\frac{\partial X^{\mu }}{%
\partial z},\frac{\partial X^{\mu }}{\partial v}})\text{ \ ,}  \tag{7.10}
\end{equation}%
where 
\begin{equation*}
K_{1}\equiv r_{c}^{2}e^{-2kr_{-}\Phi }[(\frac{\partial \Phi }{\partial z}%
)^{2}(\frac{\partial X^{1}}{\partial v})^{2}+(\frac{\partial \Phi }{\partial
v})^{2}(\frac{\partial X^{1}}{\partial z})^{2}-(\frac{\partial \Phi }{%
\partial z})^{2}\sum_{i=2}^{4}(\frac{\partial X^{i}}{\partial v})^{2}-
\end{equation*}%
\begin{equation*}
-(\frac{\partial \Phi }{\partial v})^{2}\sum_{i=2}^{4}(\frac{\partial X^{i}}{%
\partial z})^{2}]+3r_{c}^{4}(\frac{\partial \Phi }{\partial z})^{2}(\frac{%
\partial \Phi }{\partial v})^{2}-8r_{c}^{2}\frac{\partial \Phi }{\partial z}%
\frac{\partial \Phi }{\partial v}\frac{\partial X^{1}}{\partial z}\frac{%
\partial X^{1}}{\partial v}e^{-2kr_{-}\Phi }+
\end{equation*}%
\begin{equation}
+8r_{c}^{2}e^{-2kr_{-}\Phi }\frac{\partial \Phi }{\partial z}\frac{\partial
\Phi }{\partial v}\sum_{i=2}^{4}\frac{\partial X^{i}}{\partial z}\frac{%
\partial X^{i}}{\partial v}\text{ \ \ \ ,}  \tag{7.11}
\end{equation}%
\begin{equation*}
K_{2}\equiv 8\frac{\partial X^{1}}{\partial z}\frac{\partial X^{1}}{\partial
v}\sum_{i=2}^{4}\frac{\partial X^{i}}{\partial z}\frac{\partial X^{i}}{%
\partial v}-\left( \frac{\partial X^{1}}{\partial z}\right)
^{2}\sum_{i=2}^{4}\left( \frac{\partial X^{i}}{\partial v}\right)
^{2}-\left( \frac{\partial X^{1}}{\partial v}\right)
^{2}\sum_{i=2}^{4}\left( \frac{\partial X^{i}}{\partial z}\right) ^{2}-
\end{equation*}%
\begin{equation*}
-3\left( \frac{\partial X^{1}}{\partial z}\right) ^{2}\left( \frac{\partial
X^{1}}{\partial v}\right) ^{2}+\left( \sum_{i=2}^{4}\frac{\partial X^{i}}{%
\partial z}\right) ^{2}\left( \sum_{i=2}^{4}\frac{\partial X^{i}}{\partial z}%
\right) ^{2}-
\end{equation*}%
\begin{equation}
-4\left( \sum_{i=2}^{4}\frac{\partial X^{i}}{\partial z}\frac{\partial X^{i}%
}{\partial v}\right) ^{2}\text{ \ \ \ .}  \tag{7.12}
\end{equation}%
Setting up $\Phi (z,v)=\pi $ (note that then $K_{1}=\pi $), one obtains 
\begin{equation}
\sqrt{-g}_{vis}=L_{1}(\Phi ,X^{\mu },\sqrt{-g})\sqrt{-g}\text{ \ \ \ \ ,} 
\tag{7.13}
\end{equation}%
where 
\begin{equation}
L_{1}(\Phi ,X^{\mu },\sqrt{-g})\equiv \frac{e^{-2kr_{-}\pi }}{%
e^{-2kr_{-}\Phi }}\sqrt{1-\frac{K_{1}(\Phi ,X^{\mu })}{\left( \sqrt{-g}%
\right) ^{2}}\text{ \ \ .}}  \tag{7.14}
\end{equation}%
Unlike the previously discussed in [15] case, when the \textquotedblright
visible\textquotedblright\ volume element is represented as a product of
some factor (constant), multiplying the volume element $\sqrt{-g}$ (i.e. $%
\sqrt{-g}_{vis}=e^{4kr_{-}\pi }\sqrt{-g}$), the present case might seem to
be quite different, since the function $L_{1}$ depends again on $\sqrt{-g}$.
However, it shall be proved below that by requiring the action of the
\textquotedblright visible\textquotedblright\ brane to remain unchanged
after the rescaling (7.9), still such a possibility will exist, but in a
more general form. Indeed, after the rescaling (7.9) $H\rightarrow 
\widetilde{H}f$ \ \ ;\ \ $v_{0}\rightarrow \widetilde{v}_{0}$ the action
(7.2) becomes (written in the two - dimensional coordinates $(z,v)$) 
\begin{equation*}
S_{vis}=\dint dzdv\sqrt{-g}L_{3}L_{1}\left[ g^{\mu \nu }L_{2}\text{ }%
f^{2}D_{\mu }\widetilde{H}^{+}D_{\nu }\widetilde{H}-\lambda f^{4}\left( \mid 
\widetilde{H}\mid ^{2}-\widetilde{v}_{0}^{2}\right) ^{2}\right] +\ \ \ 
\end{equation*}%
\begin{equation}
+\int dzdv\sqrt{-g}\text{ }L_{3}L_{add}\text{ \ \ ,}  \tag{7.15}
\end{equation}%
where 
\begin{equation*}
L_{add}\equiv L_{2}g^{\mu \nu }[\mid \widetilde{H}\mid ^{2}A_{\nu }\partial
_{\mu }\mid f\mid ^{2}+\mid \widetilde{H}\mid ^{2}A_{\nu }\partial _{\mu
}f^{+}\partial _{\nu }f+
\end{equation*}%
\begin{equation}
+\widetilde{H}^{+}f\text{ }\partial _{\mu }f^{+}\partial _{\nu }\widetilde{H}%
+\widetilde{H}\text{ }f\text{ }^{+}\partial _{\mu }f\text{ }\partial _{\nu }%
\widetilde{H}^{+}\text{ }  \tag{7.16}
\end{equation}%
and the covariant derivative $D_{\mu }$ is expressed as $D_{\mu }=\partial
_{\mu }+A_{\mu }$. Clearly, the visible brane actions before and after the
rescaling will remain unchanged if 
\begin{equation}
L_{1}L_{2}f^{2}=1\text{ \ \ \ ; \ \ \ \ \ \ }L_{1}f^{4}=1  \tag{7.17}
\end{equation}%
and 
\begin{equation}
L_{add}=0\text{ \ \ \ \ \ .}  \tag{7.18}
\end{equation}%
The first two relations (7.17) give 
\begin{equation}
f=\pm (L_{2})^{\frac{1}{2}}=\pm (L_{1})^{-\frac{1}{6}}\text{ \ \ \ \ ,} 
\tag{7.19}
\end{equation}%
which can be rewritten as 
\begin{equation}
\frac{1}{L_{2}^{3}}=\frac{e^{-2kr_{-}\pi }}{e^{-2kr_{-}\Phi }}\sqrt{1-\frac{%
K_{1}(\Phi ,X^{\mu })}{(\sqrt{-g})^{2}}\text{ }}\text{ \ \ \ \ \ ,} 
\tag{7.20}
\end{equation}%
from where the function $K_{1}(\Phi ,X^{\mu })$ can be expressed and
substituted into expression (7.10) for $\sqrt{-g}$. Thus one obtains 
\begin{equation}
\sqrt{-g}=L_{2}^{3}e^{-2kr_{-}\pi }\sqrt{K_{2}(X^{\mu })}\text{ \ \ \ .} 
\tag{7.21}
\end{equation}%
From (7.10) for $\Phi (z,v)=\pi $ one can easily derive 
\begin{equation}
\sqrt{-g}_{vis}=\sqrt{e^{-4kr_{-}\pi }}.\sqrt{K_{2}(X^{\mu })}=\frac{1}{%
L_{2}^{3}}\sqrt{-g}\text{ \ \ .}  \tag{7.22}
\end{equation}%
Therefore, even in the more general case of contravariant metric tensor,
different from the inverse one, there is a relation similar to $\sqrt{-g}%
_{vis}=e^{-4kr_{-}\pi }\sqrt{-g}$, but with the function $\frac{1}{L_{2}^{3}}
$, multiplying the volume element. Let us remind that for performing the
calculation it was sufficient to know the function $\Phi (z,v)$ as a
solution of the system of nonlinear differential equations, but not the
points $(z_{0}^{(l)},v_{0}^{(l)})$, at which $\Phi
(z=z_{0}^{(l)},v=v_{0}^{(l)})=\pi $. Consequently, in the final result
(7.22) one cannot set up 
\begin{equation}
\sqrt{-g}_{vis}\left( \mathbf{X(}z=z_{0}^{(l)},v=v_{0}^{(l)}),\Phi =\pi
\right) =\frac{1}{L_{2}^{3}(z=z_{0}^{(l)},v=v_{0}^{(l)},\Phi =\pi )}\sqrt{-g}
\tag{7.23}
\end{equation}%
Then to any mass $m_{0}$\ on the visible three- brane would correspond a
single physical mass, \textquotedblright measured\textquotedblright\ with
the metric $g^{\mu \nu }$ 
\begin{equation}
m^{(l)}=m_{0}f=m_{0}\sqrt[4]{L_{2}^{(l)}(z=z_{0}^{(l)},v=v_{0}^{(l)},\Phi
=\pi )\text{ }}\text{ \ \ \ ,}  \tag{7.24}
\end{equation}%
i. e. there is no degeneracy of masses. The corresponding additional
condition (7.18) $L_{add}=0$ can be written as 
\begin{equation*}
L_{add}=f^{2}\text{ }\partial _{\mu }\ln f\text{ }[2\mid \widetilde{H}\mid
^{2}A_{\nu }+2\widetilde{H}^{+}+\widetilde{H}^{2}\text{ \ }\partial _{\nu
}\left( \frac{\widetilde{H}^{+}}{\widetilde{H}}\right) +
\end{equation*}%
\begin{equation}
+2\mid \widetilde{H}\mid ^{2}\partial _{\nu }(\ln f)\text{ }-\widetilde{H}%
^{2}\partial _{\nu }(\ln f)]=0\text{ \ \ \ ,}  \tag{7.25}
\end{equation}%
from where the trivial case $f=const$ is obtained from $\partial _{\mu }\ln
f=0$.

Let us now see how the above approach will change if the points $%
(z_{0}^{(l)},v_{0}^{(l)})$ on the complex plane, at which the equation $\Phi
(z=z_{0}^{(l)},v=v_{0}^{(l)})=\pi $ holds, are considered to be known. The
function $L_{2}(z,v)$ in the ratio of $g_{vis}^{\mu \nu }$ and $g^{\mu \nu }$
will be different and will be denoted as $\widetilde{L}_{2}(z,v)$ 
\begin{equation}
\widetilde{L}_{2}(z,v)\equiv \frac{F_{\mu }(\mathbf{X}%
(z_{0}^{(l)},v_{0}^{(l)}),\Phi =\pi ,z_{0}^{(l)})F_{\nu }(\mathbf{X}%
(z_{0}^{(l)},v_{0}^{(l)}),\Phi =\pi ,z_{0}^{(l)})}{F_{\mu }(\mathbf{X}%
(z,v),\Phi (z,v),z)F_{\nu }(\mathbf{X}(z,v),\Phi (z,v),z)}\text{ \ .} 
\tag{7.26}
\end{equation}%
Also from formulae (7.9) for $\Phi =\pi $ and for all points $%
(z,v)=(z_{0}^{(l)},v_{0}^{(l)})$ one can obtain 
\begin{equation}
\sqrt{-g}_{vis}=\sqrt{-g}e^{-2kr_{-}\pi }\sqrt{\frac{K_{2}^{0}(X^{\mu
}(z_{0}^{(1)},v_{0}^{(1)})}{K_{1}+e^{-4kr_{-}\Phi }K_{2}(X^{\mu }(z,v))}}=%
\widetilde{L}_{1}(z,v)\text{ \ \ \ ,}  \tag{7.27}
\end{equation}%
which evidently is different from expression (7.22). Consequently, for this
case instead of (7.20) one receives 
\begin{equation}
\frac{1}{\widetilde{L}_{2}^{6}}=e^{-4kr_{-}\pi }\frac{K_{2}^{0}(X^{\mu
}(z_{0}^{(1)},v_{0}^{(1)})}{K_{1}+e^{-4kr_{-}\Phi }K_{2}(X^{\mu }(z,v))}%
\text{ \ \ ,}  \tag{7.28}
\end{equation}%
from where the function $K_{1}$ can be expressed and substituted into
expression (7.27) for $\sqrt{-g}_{vis}$. Taking into account again equality
(7.10) for $\sqrt{-g}$, one obtains 
\begin{equation}
\sqrt{-g}_{vis}=\sqrt{-g}\frac{1}{\widetilde{L}_{2}^{3}}=\frac{\sqrt{%
K_{2}^{0}}}{e^{2kr_{-}\pi }}\text{ \ \ \ .}  \tag{7.29}
\end{equation}%
Therefore, the volume element of the \textquotedblright
visible\textquotedblright\ brane is a constant, while the real volume
element $\sqrt{-g}$ is $\widetilde{L}_{2}^{3}$ times the volume of the
\textquotedblright visible\textquotedblright\ brane.

In this case, to any mass $m_{0}$ on the ''visible'' brane would correspond $%
l$ in number physical masses, determined by the formulae 
\begin{equation}
m^{(l)}=m_{0}f^{(l)}=m_{0}\sqrt{\widetilde{L}_{2}^{(l)}(z,v)\text{ }}\text{
\ \ \ ,}  \tag{7.30}
\end{equation}
where the function $\widetilde{L}_{2}(z,v)$ is given by (7.26). Therefore,
there will be a degeneracy of masses.

\section{ CONCLUSION}

\bigskip Let us summarize the most important proposals and results in this
(first) part of the paper and give also some suggestions for future research
on the base of the refinement of some of the initial assumptions:

\bigskip 1. If the Randall - Sundrum model is investigated within the
framework of the multidimensional Lobachevsky space, then there should be
some corrections to the extradimensional volume element and to the Newton's
constant. In principle, it is known how the Newton's force law can be
derived for the $4D$ Lobachevsky space, so probably it can be extended to
more dimensions. Note that the corrections to the extradimensional volume
can be found after performing the integration in (3.10), using expression
(3.7) for the $d-$dimensional hyperbolic volume.

2. The orbifold identification of the points $-\pi r_{C}$ and $-\pi r_{C}$
under compactification is performed. Note that the choice of the
uniformization coordinate $z$ as $z=\pi r_{c}(\cos \Phi +i\sin \Phi )$ is an
approximation and need not to be done, since the dependence of the angular
coordinate $\Phi $ on $z$ should be obtained after finding the algebraic
solutions of the cubic equation and performing the integration of the system
of nonlinear differential equations, as this was pointed out in [10]. Some
particular simple choice of the metric has to be made - the metric (2.4) is
fully appropriate for that purpose.

3. Coordinate transformations (4.4) - (4.5), containing the distance $\rho $
in the Lobachevsky space have been performed with respect to the metric
(2.4). The choice $\widetilde{k}\neq \frac{1}{c}$, when the extradimensional
distance exponentially increases, seems to be not consistent with the
Lobachevsky geometry, since it is expected to go back to the euclidean
geometry in the limit $c\rightarrow \infty $. However, this is not the case
since the "absence" of such a constant $\widetilde{k}=\frac{1}{c}$ makes
such a transition impossible, and this turns out to be physically
consistent. Of course, the same approach may be applied to more complicated
models with an arbitrary \textquotedblright warp\textquotedblright\
exponential factor and $(D-4)$ compact non-flat extra- dimensional spacetime
[29, 30] 
\begin{equation}
ds^{2}=g_{ab}(\mathbf{X})dX^{a}dX^{b}=2e^{2A(y)}\eta _{\mu \nu }dX^{\mu
}dX^{\nu }+h_{ij}(y)dy^{i}dy^{j}\text{ \ \ \ ,}  \tag{8.1}
\end{equation}%
where $(a,b)=1,2,...D$; $(\mu ,\nu )=1,..,4$; $(i,j)=5,...,D$. The
transformations (4.4 - 4.5) can again be used (with $a=b=\widetilde{k}$ $%
e^{-A(y)}$) and an expression for the warp factor $A(y)$ can be found so
that the metric tensor components $h_{ij}$ of the extra- dimensional space
are left unchanged. In principle, the motivation for different warp factors
comes from $M$- theory (see [31] for a recent review),

4. The algebraic equations for the preservation of the scalar curvature
invariant in section 5 have been obtained. In the general case, for an
arbitrary tensor $\widetilde{g}^{\mu \nu }$, the exact solution of the
problem about the conservation of the scalar curvature invariant requires
the solution of the fourth-order algebraic equation (5.10) with respect to
the components of the tensor $\widetilde{g}^{\mu \nu }$, under the
fulfillment also of (5.11). On the base of the algorithm, presented in
[8-10], this can be performed, and the solution will be greatly facilitated
by the fact that there are no derivatives of $\widetilde{g}^{\mu \nu }$.

5. The two three - brane model in section 7 has been investigated before in
numerous papers, taking into account a more complicated physical setting.
Concretely, in [38] the two branes are considered as two positive tension
walls, separated by a distance, corresponding to the inverse of the GUT
scale. Therefore, the wall tension terms in the effective field theory are
taken into account. In section 7 a more simplified model has been presented,
having the purpose to set up the mathematical background for the Higgs mass
generation in such a two-brane model in more general gravitational
theories.The key moment in the problem is how many points $(z_{0},v_{0})$
satisfy the equation $\Phi (z_{0},v_{0})=\pi $. Since it is not known
whether and under what choice of the metric they can be found, evidently the
result has a qualitative character and not a quantitative one. Moreover,
again on the base of the papers [10], the initial metric in the coordinates $%
(X^{\mu }$,$\Phi )$ is mapped into a two-dimensional one with coordinates $%
(z,v)$, so this mapping yet is not studied, neither is known whether the
correspondence between the two metrics is a unique one.

\section{\protect\bigskip APPENDIX\ A: ALGEBRAIC\ \ EQUATIONS\ \ IN\ \
GRAVITY\ \ THEORY}

\bigskip In this Appendix some basic information about the algebraic
geometry approach wil be provided, which was initially developed in [8] and
subsequently in [9, 10]. This knowledge will be necessary in order to
understand how formulae (7.5) has emerged.

The algebraic geometry approach and the derivation of the basic algebraic
equations in gravity theory is based on two different representations of the
gravitational Lagrangian, which subsequently are assumed to be equal.

The standard \textit{(first) representation} of the gravitational Lagrangian
is based on the standard Christoffell connection $\Gamma _{ij}^{k}$, the
Ricci tensor $R_{ik}$ and \textit{another contravariant tensor}, chosen for
this partial case in the form of the factorized product $\widetilde{g}%
^{ij}=dX^{i}dX^{j}$ [10] 
\begin{equation}
L_{1}=-\sqrt{-g}\widetilde{g}^{ik}R_{ik}=-\sqrt{-g}dX^{i}dX^{k}R_{ik}\text{
\ .}  \tag{A1}
\end{equation}%
The choice of this (another) contravariant tensor, which is not the inverse
one to the covariant one, is motivated by the affine geometry approach and
the gravitational theories with covariant and contravariant metrics and
connections, the essence of which was clarified in the introduction of this
paper.

In the \textit{second representation, }the\textit{\ }Christoffell connection 
$\widetilde{\Gamma }_{ij}^{k}$ and the Ricci tensor $\widetilde{R}_{ik}$%
\textit{\ }are the "tilda" quantities 
\begin{equation}
\widetilde{R}_{ij}=\widetilde{R}_{ji}=\partial _{k}\widetilde{\Gamma }%
_{ij}^{k}-\partial _{i}\widetilde{\Gamma }_{kj}^{k}+\widetilde{\Gamma }%
_{kl}^{k}\widetilde{\Gamma }_{ij}^{l}-\widetilde{\Gamma }_{ki}^{m}\widetilde{%
\Gamma }_{jm}^{k}\text{ \ ,}  \tag{A2}
\end{equation}
meaning that the "tilda" Christoffell connection is determined by formulae 
\begin{equation}
\widetilde{\Gamma }_{kl}^{s}\equiv \frac{1}{2}%
dX^{i}dX^{s}(g_{ik,l}+g_{il,k}-g_{kl,i})  \tag{A3}
\end{equation}
with the new contravariant tensor $\widetilde{g}^{ij}=dX^{i}dX^{j}$. Thus
the expression for the \textit{second representation} of the gravitational
Lagrangian acquires the form 
\begin{equation*}
L_{2}=-\sqrt{-g}\widetilde{g}^{il}\widetilde{R}_{il}=
\end{equation*}%
\begin{equation}
=-\sqrt{-g}dX^{i}dX^{l}\{p\Gamma _{il}^{r}g_{kr}dX^{k}-\Gamma
_{ik}^{r}g_{lr}d^{2}X^{k}-\Gamma _{l(i}^{r}g_{k)r}d^{2}X^{k}\}\text{ .} 
\tag{A4}
\end{equation}
The condition for the \textit{equivalence of the two representations} $%
L_{1}=L_{2}$ gives a cubic algebraic equation with respect to the algebraic
variety of the first differential $dX^{i}$ and the second ones $d^{2}X^{i}$
[10] 
\begin{equation}
dX^{i}dX^{l}\left( p\Gamma _{il}^{r}g_{kr}dX^{k}-\Gamma
_{ik}^{r}g_{lr}d^{2}X^{k}-\Gamma _{l(i}^{r}g_{k)r}d^{2}X^{k}\right)
-dX^{i}dX^{l}R_{il}=0\text{ \ \ \ \ , }  \tag{A5}
\end{equation}

where $p$ is the scalar quantity

\begin{equation}
p\equiv div(dX)\equiv \frac{\partial (dX^{l})}{\partial x^{l}}\text{,} 
\tag{A6}
\end{equation}%
which \textquotedblright measures\textquotedblright\ the divergency\textbf{\ 
}of the vector field $dX$. The algebraic variety of the algebraic equation
(A5) (i.e. the set of variables, with respect to which the equation is
solved and which, if substituted, satisfy it) consists of the differentials $%
dX^{i\text{ }}$ and their derivatives $\frac{\partial (dX^{s})}{\partial
x^{k}}$.\bigskip

Similarly, in [9] it was proved that a cubic algebraic equation for
reparametrization invariance of the gravitational Lagrangian 
\begin{equation*}
\widetilde{g}^{i[k}\widetilde{g}_{,l}^{l]s}\Gamma _{ik}^{r}g_{rs}+\widetilde{%
g}^{i[k}\widetilde{g}^{l]s}\left( \Gamma _{ik}^{r}g_{rs}\right) _{,l}+
\end{equation*}%
\begin{equation}
+\widetilde{g}^{ik}\widetilde{g}^{ls}\widetilde{g}^{mr}g_{pr}g_{qs}\left(
\Gamma _{ik}^{q}\Gamma _{lm}^{p}-\Gamma _{il}^{p}\Gamma _{km}^{q}\right) -R=0%
\text{ \ \ \ \ }  \tag{A7}
\end{equation}%
can be obtained also in the case of a generally chosen contravariant metric
tensor components (for which $\widetilde{g}^{ij}\neq dX^{i}dX^{j}$). Also,
the Einstein's system of equations can also be written in the form of a
system of cubic algebraic equations [9] with respect to the contravariant
components, but this is irrelevant to the investigation in this paper.

Further, in [9, 10] the solutions of the algebraic equation (A5) have been
found. The main peculiarity of the proposed new method for finding the
solutions for the contravariant metric components are the following:

1. They are found for the particular case $g_{\alpha \mu }\widetilde{g}^{\mu
\nu }=l_{\alpha }^{\nu }\neq $ $\delta _{\alpha }^{\nu }$, when the
contravariant components are not inverse ones to the covariant components.
In fact, the algebraic equation (A5) is valid for such a case and under the
additional restriction $\widetilde{g}^{ij}=dX^{i}dX^{j}$.

2. It has been assumed also that $d^{2}X^{i}=0$.

3. The algebraic equation (A5) is a \textit{multivariable cubic algebraic
equation} (since the contravariant components $\widetilde{g}^{\mu \nu }$ in
the general $n-$dimensional case are $\frac{n(n-1)}{2}$ in number), which is
a substantial difference from the two-dimensional case.

For the two-dimensional case, it is known how to parametrize the following
two dimensional cubic algebraic equation%
\begin{equation}
y^{2}=4x^{3}-g_{2}x-g_{3}\text{ \ \ \ ,}  \tag{A8}
\end{equation}%
where $g_{2}$ and $g_{3}$ are the complex numbers, called the \textit{%
Eisenstein series} 
\begin{equation}
g_{2}=60\sum\limits_{\omega \subset \Gamma }\frac{1}{\omega ^{4}}\text{ ; }\
\ \text{\ }g_{3}=140\sum\limits_{\omega \subset \Gamma }\frac{1}{\omega ^{6}}%
\text{ \ \ .}  \tag{A9}
\end{equation}%
The basic and very simple idea about parametrization of the cubic algebraic
equation (A8) with the Weierstrass function (see the monograph [39] for an
excellent intoduction into this problem) can be presented as follows: Let us
define the lattice $\Lambda =\{m\omega _{1}+n\omega _{2}\mid m,n\in Z;$ $%
\omega _{1},\omega _{2}\in C,Im\frac{\omega _{1}}{\omega _{2}}>0\}$ and the
mapping $f:$ $C/\Lambda \rightarrow CP^{2}$, which maps the factorized
(along the points of the lattice $\Lambda $) part of the points on the
complex plane into the two dimensional complex projective space\textbf{\ }$%
CP^{2}$. This means that each point $z$ on the complex plane is mapped into
the point $(x,y)=(\rho (z),\rho ^{^{\prime }}(z))$, where $x$ and $y$ belong
to the affine curve (A8) and $\rho (z)$ denotes the \textit{Weierstrass
elliptic function } 
\begin{equation}
\rho (z)=\frac{1}{z^{2}}+\sum\limits_{\omega }\left[ \frac{1}{(z-\omega )^{2}%
}-\frac{1}{\omega ^{2}}\right]  \tag{A10}
\end{equation}%
and the summation is over the poles in the complex plane. In other words,
the functions $x=\rho (z)$ and $y=\rho ^{^{\prime }}(z)$ are uniformization
functions for the cubic curve \ and it can be proved [39] that \textit{the
only cubic algebraic curve with number coefficients} which is parametrized
by the uniformization functions $x=\rho (z)$ and $y=\rho ^{^{\prime }}(z)$
is the affine curve (A8).

In order to parametrize the multivariable cubic algebraic equation (A5),
again the parametrization of the two-dimensional equation (A8) has to be
used. For the purpose, the approach of the s.c.\textit{"embedding sequence
of cubic algebraic equations"} has been introduced in [9, 10], the essence
of which in brief is the following:

The initial cubic multivariable algebraic equation (A5) is presented as a
cubic equation with respect to the variable $dX^{3}$ only (for simplicity,
the three-dimensional case is taken, but the approach can be generalized to
any dimensions) 
\begin{equation}
A_{3}(dX^{3})^{3}+B_{3}(dX^{3})^{2}+C_{3}(dX^{3})+G^{(2)}(dX^{2},dX^{1},g_{ij},\Gamma _{ij}^{k},R_{ik})=0%
\text{ \ \ ,}  \tag{A11}
\end{equation}%
where the coefficient functions $A_{3}$, $B_{3}$ , $C_{3}$ and $G^{(2)}$
depend on the variables $dX^{1}$ and $dX^{2}$ of the algebraic subvariety
and on the metric tensor $g_{ij}$, the Christoffel connection $\Gamma
_{ij}^{k}$ and the Ricci tensor $R_{ij}$. Further the Greek indices $\alpha
,\beta $ take values $\alpha ,\beta =1,2$ while the indice $r$ takes values $%
r=1,2,3$.

In order to obtain the embedded sequence of equations, a linear-fractional
transformation 
\begin{equation}
dX^{3}=\frac{a_{3}(z)\widetilde{dX}^{3}+b_{3}(z)}{c_{3}(z)\widetilde{dX}%
^{3}+d_{3}(z)}  \tag{A12}
\end{equation}%
is performed with the purpose of setting up to zero the coefficient
functions in front of the highest (third) degree of $\ \widetilde{dX}^{3}$
in the newly derived (i.e. transformed) cubic equation. This will be
achieved if $G^{(2)}(dX^{2},dX^{1},g_{ij},\Gamma _{ij}^{k},R_{ik})=-\frac{%
a_{3}Q}{c_{3}^{3}}$, which can be rewritten in the form of a two-dimensional
cubic algebraic equation with respect to the\textbf{\ }algebraic variety of
the variables $dX^{1}$ and $dX^{2}$: 
\begin{equation}
p\Gamma _{\gamma (\alpha }^{r}g_{\beta )r}dX^{\gamma }dX^{\alpha }dX^{\beta
}+K_{\alpha \beta }^{(1)}dX^{\alpha }dX^{\beta }+K_{\alpha }^{(2)}dX^{\alpha
}+2p\left( \frac{a_{3}}{c_{3}}\right) ^{3}\Gamma _{33}^{r}g_{3r}=0  \tag{A13}
\end{equation}
\ \ \ and $K_{\alpha \beta }^{(1)}$ and $K_{\alpha }^{(2)}$ again depend on $%
R_{\alpha \beta }$, $\Gamma _{\alpha \beta }^{r}$, $g_{\beta r}$ and the
ratios $\frac{a_{3}}{c_{3}}$ and $\frac{d_{3}}{c_{3}}$. The originally given
equation (A5) is called \textit{"the embedding equation"} for the equation
(A13). Consequently, in the general case of an $n-$dimensional cubic
equation, after applying the described above algorithm, one would obtain an $%
(n-1)-$ dimensional cubic equation, afterwards again - an $(n-2)-$%
dimensional equation and so on. \textit{In other words, this is the s.c.
"embedding sequence" of cubic algebraic equations.}

In the case of the "transformed" two-dimensional equation (A5) (with respect
to the variables $n_{3}=\widetilde{dX}^{3}$ and $m$ $=\frac{a_{3}}{c_{3}}$),
in [8] it has been proved how it can be brought to an equation of the kind 
\begin{equation}
\widetilde{n}^{2}=\overline{P}_{1}(\widetilde{n})\text{ }m^{3}+\overline{P}%
_{2}(\widetilde{n})\text{ }m^{2}+\overline{P}_{3}(\widetilde{n})\text{ }m+%
\overline{P}_{4}(\widetilde{n})\text{ ,}  \tag{A14}
\end{equation}

where $\overline{P}_{1}(\widetilde{n})$ $,\overline{P}_{2}(\widetilde{n}),$ $%
\overline{P}_{3}(\widetilde{n})$, $\overline{P}_{4}(\widetilde{n})$ are
complicated functions of the ratios $\frac{c_{3}}{d_{3}}$, $\frac{b_{3}}{%
d_{3}}$ and $A_{3},B_{3},C_{3}$ and the variable $\widetilde{n}$ is related
to the variable $n$ through a definite linear transformation. From (A14),
one can obtain the parametrizable form 
\begin{equation}
\widetilde{n}^{2}=4m^{3}-g_{2}m-g_{3}\text{ }  \tag{A15 }
\end{equation}

of the cubic algebraic equation, from where the solution for $dX^{3}$ can be
expressed as 
\begin{equation}
dX^{3}=\frac{\frac{b_{3}}{c_{3}}+\frac{\rho (z)\rho ^{^{\prime }}(z)}{\sqrt{%
k_{3}}\sqrt{C_{3}}}-L_{1}^{(3)}\frac{B_{3}}{C_{3}}\rho (z)-L_{2}^{(3)}\rho
(z)}{\frac{d_{3}}{c_{3}}+\frac{\rho ^{^{\prime }}(z)}{\sqrt{k_{3}}\sqrt{C_{3}%
}}-L_{1}^{(3)}\frac{B_{3}}{C_{3}}-L_{2}^{(3)}}\text{ \ \ \ \ \ .}  \tag{A16}
\end{equation}
It is important to mention that in (A16) $B_{3}$ and $C_{3}$ are complicated
functions, depending on $dX^{1}$and $dX^{2}$, due to which (A16) can be
called the \textit{"embedding solution"} for $dX^{1}$ and $dX^{2}$.

After applying the same parametrization procedure with respect to the
embedded equations, one can obtain a similar expression for $dX^{2}$ as an
embedding solution for $dX^{1}$ and an expression for $dX^{1}$.
Consequently, all the solutions (for $l=1,2,3$) can be written as

\begin{equation}
dX^{l}(X^{1},X^{2},X^{3})=F_{l}(g_{ij}(\mathbf{X}),\Gamma _{ij}^{k}(\mathbf{X%
}),\rho (z),\rho ^{^{\prime }}(z))=F_{l}(\mathbf{X},z)\text{ \ \ \ ,} 
\tag{A17}
\end{equation}
and the functions $F_{l}(\mathbf{X},z)$ are \textit{"parametrization"
functions} for the initially given algebraic equation (A5). \textit{However,
yet it is not justified to call them "uniformization functions", since they
depend not only on the complex variable }$z$\textit{, but also on the
generalized coordinates }$X$\textit{.}

Now it shall be proved that these functions can be considered also as 
\textit{"uniformization functions"}. But as a first step, one should
reconcile the appearence of the additional complex coordinate $z$ on the
right-hand side of \ (A17) with the dependence of the differentials on the
left-hand side of (A17) only on the generalized coordinates $%
(X^{1},X^{2},X^{3})$ (and on the initial coordinates $x^{1},x^{2},x^{3}$
because of the mapping $X^{i}=X^{i}(x^{1},x^{2},x^{3})$). The only
reasonable assumption will be that \textit{the initial coordinates depend
also on the complex coordinate}, i.e. 
\begin{equation}
X^{l}\equiv X^{l}(x^{1}(z),x^{2}(z),x^{3}(z))=X^{l}(\mathbf{x,}\text{ }z)%
\text{ \ \ \ \ .}  \tag{A18}
\end{equation}

Further, the important initial assumptions ($l=1,2,3$) 
\begin{equation}
d^{2}X^{l}=0=dF_{l}(\mathbf{X}(z),z)=\frac{dF_{l}}{dz}dz\text{ \ \ ,} 
\tag{A19}
\end{equation}%
should be taken into account, from where one easily gets the system of 
\textit{three inhomogeneous linear algebraic equations} with respect to the
functions $\frac{\partial X^{1}}{\partial z}$, $\frac{\partial X^{2}}{%
\partial z}$ and $\frac{\partial X^{3}}{\partial z}$ ($l=1,2,3$) 
\begin{equation}
\frac{\partial F_{l}}{\partial X^{1}}\frac{\partial X^{1}}{\partial z}+\frac{%
\partial F_{l}}{\partial X^{2}}\frac{\partial X^{2}}{\partial z}+\frac{%
\partial F_{l}}{\partial X^{3}}\frac{\partial X^{3}}{\partial z}+\frac{%
\partial F_{l}}{\partial z}=0\text{ \ \ \ .}  \tag{A20}
\end{equation}%
The solution of this algebraic system ($i,k,l=1,2,3$) 
\begin{equation}
\frac{\partial X^{l}}{\partial z}=G_{l}\left( \frac{\partial F_{i}}{\partial
X^{k}}\right) =G_{l}\left( X^{1},X^{2},X^{3},z\right) \text{ \ \ \ \ \ } 
\tag{A21}
\end{equation}%
represents a system of \textit{three first - order nonlinear differential
equations}.\textbf{\ }A solution of this system can always be found in the
form \textbf{\ } 
\begin{equation}
X^{1}=X^{1}(z)\text{ \ \ ; \ \ \ }X^{2}=X^{2}(z)\text{ \ \ ; \ \ \ \ }%
X^{3}=X^{3}(z)\text{ \ \ \ \ \ \ \ \ \ .}  \tag{A22}
\end{equation}%
and therefore, the metric tensor components will also depend \textit{only }
on the complex coordinate $z$, i.e. $g_{ij}=g_{ij}(\mathbf{X}(z))$. Thus it
is proved that the functions $F_{l}(\mathbf{X},z)$ in (A17) can be
considered also to be \textit{"uniformization" functions, which depend only
on the complex variable }$z$\textit{. }

The parametrization (uniformization) of the initially given cubic algebraic
equation can \ be extended to a parametrization by means of a pair of
complex coordinates $(z,v)$\ in the following way 
\begin{equation}
dX^{i}(\mathbf{X})=F_{i}(\mathbf{X}(\mathbf{x}(z,v)),z)\text{ \ \ .} 
\tag{A23}
\end{equation}%
In [9,10] the corresponding system of equations, related to the initial
assumption $d^{2}X^{l}=0$ \ has been analysed and it has been proved that
this system is noncontradictory. Therefore, formulae (7.5) will be valid.

\section*{\protect\bigskip Acknowledgments}

\bigskip This paper is written in memory of Prof. Nikolai Alexandrovich
Chernikov (16. 12.1928 - 17.04.2007) (BLTP, JINR, Dubna), to whom I am
indebted for my understanding of non-euclidean (Lobachevsky) geometry.

The author is very grateful to Prof. V. V. Nesterenko (BLTP, JINR, Dubna)
and to Dr. O. P. Santillan (IAFE, Buenos Aires), Dr. N. S. Shavokhina (BLTP
\& LNP,JINR, Dubna) for valuable comments, discussions and critical remarks.

\ The author is grateful also to Dr.C. Kokorelis (Institute for Nuclear \&
Particle Physics N. C. S. R. Demokritos, Athens, Greece) and Dr. Al. Krause
(ASC, Munich) for bringing some references to my attention.

\end{document}